\documentclass[%
 reprint,
 amsmath,amssymb,
 aps,
]{revtex4-2}

\usepackage{graphicx}
\usepackage{dcolumn}
\usepackage{bm}
\usepackage{xcolor}
\usepackage[backgroundcolor=green,linecolor=green]{todonotes} 

\usepackage{subcaption}
\usepackage{currvita}
\begin{document}
\preprint{APS/123-QED}

\title{Bosonization of Majorana modes and edge states}

\author{Arkadiusz Bochniak}
\email{arkadiusz.bochniak@doctoral.uj.edu.pl}
\author{B\l{}a\.z{}ej Ruba}%
\email{blazej.ruba@doctoral.uj.edu.pl}%
\author{Jacek Wosiek}
\email{jacek.wosiek@uj.edu.pl}
\affiliation{%
 Institute of Theoretical Physics, Jagiellonian University, Poland.
}%

\date{\today}

\begin{abstract}
We present a bosonization procedure which replaces fermions with generalized spin variables subject to local constraints. It~requires that the number of Majorana modes per lattice site matches the coordination number modulo two. If this condition is not obeyed, then bosonization introduces additional fermionic excitations not present in the original model. In the case of one Majorana mode per site on a~honeycomb lattice, we recover a sector of Kitaev's model. We discuss also decagonal and rectangular geometries and present bosonization of the Hubbard model. For geometries with a boundary we find that certain fermionic edge modes naturally emerge. They are of different nature than edge modes encountered in topological phases of matter. Euclidean representation for the unconstrained version of a spin system of the type arising in our construction is derived and briefly studied by computing some exact averages for small volumes.
\end{abstract}

\maketitle


\section{Introduction}
\label{sec:intro}

Bosonization is an old subject, which is of interests both in condensed matter and high energy physics. The Jordan-Wigner transformation \cite{JordanWigner28} is one of the most famous methods. It provides an effective bosonization procedure in $(1+1)$-dimensions. Its non-local character in higher dimensions leads to the search of alternative methods. There exists a zoo of proposals \cite{CHEN2018234,ChenKapustin19, Wosiek:1981mn, BR2020,BURGESS199418,BRAVYI2002210,Ball05,Verstraete_2005,Fradkin89,KarchTong,Zohar18,Chen2020,Son15,ChenSon}, including approaches motivated by the Tomonaga-Luttinger model \cite{Lieb65}, generalizations or modifications of Witten's non-abelian bosonization \cite{Witten:1983ar}, as well as purely algebraic approaches \cite{Cobanera11}. Bosonization is also closely related to the subject of dualities, such as the Kramers-Wannier duality \cite{KWDuality,Wegner} or the more recent web of dualities \cite{KarchTong,KTT19,SEIBERG2016395,SENTHIL20191}.


Besides classical applications such as solving certain many body quantum models exactly \cite{SML} or overcoming sign problems in Monte Carlo studies \cite{Delgado14,Gattringer15}, bosonization has been invoked in the study of problems in quantum computation \cite{KITAEV2003,kitaev2009} and topological phases of matter \cite{gaiotto,mazaheri,KT17,Cappelli2017,Pozo21}. In \cite{Kitaev06} a two-dimensional quantum spin liquid model integrable using bosonization methods has been proposed. More recently, certain bosonization techniques were used to study inhomogeneous Luttinger liquids \cite{Huber20}, quantum phase diagram in one-dimensional superconductors \cite{Bortolin19} and also the fractional quantum Hall fluids \cite{Yang21}.

In \cite{Wosiek:1981mn} a bosonization technique, here referred to as the $\Gamma$ model, was proposed. It transforms in a local way a fermionic model into a generalized spin system subject to constraints. This correspondence was then made more precise in \cite{Szczerba:1984ca}. Generalization and new proofs were given in \cite{BR2020,BRWW2020}. Constraints present in the $\Gamma$ model were interpreted as the pure gauge condition for a certain $\mathbb{Z}_2$ gauge field. Modification of these constraints turned out to be equivalent to coupling fermions to an external gauge field. 

The most general version of the $\Gamma$ model developed so far is subject to several important limitations. First, it corresponds to a system with one fermion (hence two states) per lattice site. In this work we lift this restriction and bosonize systems with arbitrary, not necessarily even, number of Majorana modes (which are in a certain sense halves of an ordinary fermion) per lattice site. Second, in the formulations considered before the present work it was crucial that all lattice sites have an even number of neighbours. This covers many interesting examples, inlcuding the toroidal geometries frequently used in lattice simulations. Nevertheless, already for finite square lattices (say, with open boundary conditions) there exists an issue related to the existence of the boundaries. As remarked in \cite[Appendix~B]{BR2020}, coordination number changes for vertices on the boundary, which may call for an adjustment of the bosonization procedure. It was argued that some Majorana modes may be present on the boundary. In this paper we come back to this issue and discuss it in detail. 

We emphasize that the notion of a Majorana fermion used here has almost nothing to do with the one from high energy physics \cite{majorana}, which refers to a spinor field invariant under charge conjugation transformation. In particular Lorentz symmetry (or lack thereof) plays no role. Here Majorana fermions are self-adjoint operators obeying canonical anticommutation relations. Every standard fermion may be decomposed into a pair of Majoranas (real and imaginary part) in a canonical way; on the other hand pairing of Majoranas into usual fermions depends on a choice of additional structure in the space of Majorana modes \footnote{More precisely, complex structure compatible with the bilinear form determining the canonical anticommutation relations.}.

One source of interest in Majorana modes in physics comes from the theory of superconductivity \cite{schrieffer, BCS}. In the presence of Abrikosov vortices \cite{Abrikosov:1956sx} there may exist a finite number of Majorana zero modes per vortex, which resemble properties of Majorana particles \cite{WilczekMajorana}. Such phenomenon exists for eample for chiral two-dimensional $p$-wave superconductors \cite{ReadGreen}. Analogous vortex-related modes can be also found in superfluid ${}^3\mathrm{He}$ \cite{Kopnin}. Majorana zero modes are expected to appear also in the Moore–Read quantum Hall state \cite{MOORE1991362} with filling fraction $\nu=\frac{5}{2}$ (the so-called Pfaffian state). Quite generally, free fermion systems characterized by nonzero $\mathbb Z_2$ topological invariant, such as the Kitaev's quantum wires \cite{Kitaev_2001}, are expected to host Majorana zero modes on the boundary. Another exciting features of Majorana modes is their potential in topological quantum computation \cite{KITAEV2003,sarma}, related to the possibility to realize non-abelian anyons. Feasibility of such topological quantum computation is still being investigated \cite{PhysRevLett.126.090502}. 

The main idea underlying the $\Gamma$ model is to construct a representation of the even subalgebra of fermionic operators (i.e.\ the subalgebra generated by all bilinears) in terms of ``spins'' of sufficiently high dimension, or more precisely in terms of Euclidean $\Gamma$ matrices (satisfying anticommutations relations on-site, but otherwise commuting). As observed in \cite{Wosiek:1981mn}, hopping operators for fermions may be constructed given one $\Gamma(x,e)$ matrix per lattice site $x$ for every edge $e$ incident to the given site. In addition, one has to impose a certain constraints on states on the spin side. To represent the standard algebra of fermions one has to specify, besides hopping operators, also the fermionic parity operator on each site $x$. This operator has to square to $1$ and anticommute with hopping operators along all edges incident to $x$. In other words, one needs an additional $\Gamma$ matrix. If $x$ has an even number of neighbours, this additional $\Gamma$ matrix may be obtained (up to a trivial phase factor) simply by taking the product of all $\Gamma(x,e)$ with fixed $x$. We emphasize that this construction does not work if $x$ has an odd number of neighbours, since then the product of all $\Gamma(x,e)$ commutes, rather than anticommutes with individual $\Gamma(x,e)$. On the other hand introducing the additional $\Gamma$ matrix as an independent object would lead to spurious degrees of freedom.  Hence in this version one restricts to even coordination numbers.

Generalization presented in this paper is based on a few simple observations. First, in a system featuring an odd number of Majorana modes per lattice site the on-site parity operators do not exist. Therefore the $\Gamma$ model on a lattice with sites of odd degree should feature unpaired Majorana modes. Second, in presence of multiple fermionic modes per site (say, due to spin or orbital degeneracy) there exist additional independent bilinear operators which can still be bosonized if one further increases the number of $\Gamma$ matrices per lattice site. In this way one obtains a mapping between fermions and spins with only one requirement: the number of Majorana modes per site $x$ should be congruent modulo two to the number of neighbours of $x$. Even this condition can be eventually lifted. Indeed, if it is not satisfied, one can identify operators corresponding to spurious degrees of freedom and choose for them trivial dynamics decoupled from the rest of the system. 

It turns out that the $\Gamma$ model, in particular its version for arbitrary lattices proposed in this paper, shares some features with models considered in \cite{NS2009, Yao2009} and \cite{Wu2009}, despite the fact that its origin and motivation were different. We will now present a short comparison between these models. In \cite{NS2009} the vector exchange model defined in terms of bond algebra was proposed. Similarly like in our case and the models discussed by Kapustin et al. \cite{CHEN2018234,ChenKapustin19}, the main idea was to say that two models (i.e. the fermionic and bosonic ones) are equivalent if and only if their operator algebras are isomorphic. That is, the statement was purely kinematic and Hamiltonian-independent. The idea of using higher dimensional representations of Clifford algebras instead of the Pauli matrices was introduced in \cite{NS2009} in order to define higher spin (e.g. $\frac{3}{2}$, $\frac{7}{2}$ etc.) analogues of the Kitaev's model. To proceed with such fermionization procedure the need for lattices of coordination number different than $3$ emerged. The interplay between the dimension of the representation and the valency of lattice vertices is also discussed therein. Relation between constraints and the choice of a $\mathbb{Z}_2$ gauge field is also discussed and the counting of degrees of freedom is performed. In contrast, we start with the fermionic theory and perform the bosonization procedure based on the modification of the original $\Gamma$ model. The (sector of) higher spin Kitaev's model is a result of this procedure. We also allow for multiple Majorana modes on different sites and this number may in principle vary from site to site. As a consequence of the general bosonization procedure, the relation of ``unpaired" Majorana modes and generators of the Clifford algebra associated to vertices is established. The fermionization method analogous to the one in \cite{NS2009} was also proposed, at the same time, in \cite{Yao2009} and \cite{Wu2009}. In the former case the periodic boundary conditions were assumed, so that the role of the analogues of Polyakov lines discussed also in details in \cite{BRWW2020} began to be important. The role of constraints was discussed, together with the flux-attachment mechanism  \cite{Wilczek} and the interpretation of modifying the constraints as a coupling to some external $\mathbb{Z}_2$ fields. We also remark that it was argued in \cite{Yao2009} that models with a $\mathbb{Z}_2$ gauge field chosen as in \cite[Appendix~B]{BR2020} may play a role for $p$-wave superconductors. The discussion at the beginning of \cite{Wu2009} is in the same spirit as in \cite{Yao2009}. In \cite{Wu2009} the bulk-boundary correspondence is discussed in more detail for such models. As pointed out in \cite{NS2009,Yao2009,Wu2009}, these so-called $\Gamma$-matrix models may have applications for spin liquids, $(3+1)$-dimensional topological insulators and the $B$-phase of ${}^3\mathrm{He}$.

The organization of the paper is as follows. Details of our construction are presented in Section \ref{sec:1}. Then in Section \ref{sec:example} we present examples: relation to Kitaev's model on hexagonal lattice, bosonization on a decagonal lattice and bosonization of the Hubbard model on a rectangular lattice. Afterward, in Section \ref{boundary}, we discuss boundary effects in the $\Gamma$ model. We describe the example of square lattice with open boundary conditions and compare edge modes identified there with those arising on the boundary of some topological phases of matter. 

In Section \ref{Euclid} an Euclidean representation of the simplest, unconstrained $\Gamma$ model on a regular honeycomb lattice is proposed and briefly studied. 
The time evolution generated by spin Hamiltonians considered here is more complicated than in the standard Ising-like cases. Accordingly, Euclidean three-dimensional spin systems emerging in this Section are unknown and interesting by themselves. The feasibility of the standard, intermediate-volume, Monte Carlo studies is crudely assessed on the basis of the exact small-volume calculations.

\section{\label{sec:1}The bosonization method}

We consider a lattice system with fermionic degrees of freedom, whose number may vary from site to site. Real (Majorana) fermionic operators on the lattice site $x$ will be denoted by $\psi_{\alpha}(x)$, with the index $\alpha$ (labeling Majorana modes) running from $0$ to $n(x)$ with $n(x) \geq 0$. They are hermitian and satisfy anticommutation relations
\begin{equation}
\psi_{\alpha}(x) \psi_{\beta}(y) + \psi_{\beta}(y) \psi_{\alpha}(x) = 2\delta_{x,y} \delta_{\alpha, \beta}.
\end{equation}
The total number of independent Majorana operators 
\begin{equation}
n = \sum_x \left( n(x) +1 \right)
\end{equation}
is assumed to be even, which guarantees that
\begin{equation}
    (-1)^F = i^{\frac{n}{2}} \prod_{x} \prod_{\alpha=0}^{n(x)} \psi_{\alpha}(x)
\end{equation}
anticommutes with every fermion. We assume that $(-1)^F$ is an exactly conserved quantity. Otherwise the Hamiltonian may be arbitrary.

We will be interested in the algebra of even operators (i.e.\ operators commuting with $(-1)^F$). Any even operator may be expressed as a linear combination of products of bilinears of the following two types:
\begin{subequations}
\begin{align}
S(e) &= \psi_0(x) \psi_0(y) \text{ for an edge } e \text{ from } x \text{ to } y, \\
T_{\alpha}(x) &= \psi_0(x) \psi_{\alpha}(x) \text{ for } \alpha \neq 0. 
\end{align}
\label{bosonST}
\end{subequations}
All $S$ and $T$ operators are skew-hermitian and square to $-1$. Furthermore we have that
\begin{itemize}
    \item $S(e)S(e') = \pm S(e')S(e)$, with the minus sign only if $e$ shares exactly one endpoint with $e'$,
    \item $S(e)T_{\alpha}(x) = \pm  T_{\alpha}(x)S(e)$, with the minus sign only if $x$ is incident to $e$,
    \item $T_{\alpha}(x)T_{\beta}(y) = \pm T_{\beta}(y) T_{\alpha}(x)$, with the minus sign only if $x=y$ and $\alpha \neq \beta$.
\end{itemize}
It can be shown \cite{BR2020} that all relations in the algebra of even operators are generated by those given above and what will be called loop relations: if edges $e_1, \ldots, e_m$ form a loop (i.e.\ $e_i$ terminates at the initial point of $e_{i+1}$, with the convention that $e_{m+1}=e_1$), then
\begin{equation}
S(e_1) \ldots S(e_m) = 1.
\end{equation}

To bosonize the system, we generalize the approach proposed in \cite{BR2020}. For each lattice site $x$ we construct a~Clifford algebra with generators $\Gamma(x,e)$, one for each edge $e$ incident to $x$, and $\Gamma'_{\alpha}(x)$ with $0 < \alpha \leq n(x)$. They are hermitian matrices satisfying
\begin{subequations}
\begin{align}
\Gamma(x,e)\Gamma(x,e')+ \Gamma(x,e')\Gamma(x,e) &= 2 \delta_{e,e'}, \\
\Gamma'_{\alpha}(x)\Gamma'_{\beta}(x)+ \Gamma'_{\beta}(x)\Gamma'_{\alpha}(x) &= 2 \delta_{\alpha, \beta}, \\
\Gamma(x,e)\Gamma'_{\alpha}(x)+ \Gamma'_{\alpha}(x)\Gamma(x,e) &= 0.
\end{align}
\end{subequations}
Gamma matrices located on distinct lattice sites are taken to commute, and the full Hilbert space is the tensor product of on-site Hilbert spaces. In this sense the new system is bosonic. Fermionic bilinears are mapped to bosonic operators according to the local prescription
\begin{subequations}
\begin{gather}
\widehat S(e) = i \, \Gamma(x,e)\Gamma(y,e) \text{ for an edge } e \text{ from } x \text{ to } y, \\
\widehat T_{\alpha}(x) = i \, \Gamma'_{\alpha}(x) \text{ for } \alpha \neq 0, 
\end{gather}
\label{eq:bosonization}
\end{subequations}
where the hat serves as an indicator that we are referring to the bosonized operators, rather than those in the original fermionic system. It is straightforward to check that $\widehat S$ and $\widehat T$ operators satisfy all relations obeyed by $S$ and $T$, except for the loop relations. Instead, for a loop $\ell$ formed by edges $e_1,\ldots,e_m$ one has that the operator
\begin{equation}
W(\ell) = \widehat{S}(e_1) \ldots \widehat{S}(e_m)
\end{equation}
squares to $1$ and commutes with all $\widehat S$ and $\widehat T$. We are forced to impose the constraint
\begin{equation}
W(\ell) | \mathrm{phys} \rangle = | \mathrm{phys} \rangle \text{ for every loop } \ell.
\label{eq:constraints}
\end{equation}
We remark that modifying the constraint to the form
\begin{equation}
W(\ell) | \mathrm{phys} \rangle = \omega(\ell) | \mathrm{phys} \rangle
\end{equation}
with prescribed $\omega(\ell) = \pm 1$ is equivalent \cite{BR2020} to coupling fermions to a background $\mathbb Z_2$ gauge field for the $(-1)^F$ symmetry, such that $\omega(\ell)$ is the holonomy along $\ell$.

Now let $\deg(x)$ be the number of neighbors of a site $x$ and put $N(x) = \deg(x) + n(x)$. We consider the operator
\begin{equation}
\gamma(x) = i^{\frac{N(x) (N(x)-1)}{2}} \prod_{e} \Gamma(x,e) \prod_{ \alpha \neq 0} \Gamma'_{\alpha}(x).
\label{eq:gamma}
\end{equation}
Its phase is chosen so that $\gamma(x)^2=1$. If $N(x)$ is odd, $\gamma(x)$ commutes with all gamma matrices, so one may impose a relation $\gamma(x)=1$ or $\gamma(x)=-1$. This amounts to choosing one of two irreducible representation of the Clifford algebra on~$x$. If $N(x)$ is even, $\gamma(x)$ anticommutes with all gamma matrices, so it defines an additional gamma matrix: $\gamma(x) = \Gamma'_{n(x)+1}(x)$. In this case Eqs.\ \eqref{eq:bosonization} provide a bosonization of a system featuring one more Majorana fermion on the site $x$ than we have had originally. Therefore formally we bosonize only systems with all $N(x)$ odd, but the case in which this condition is not satisfied may be handled by choosing for the spurious fermions a trivially gapped Hamiltonian, not interacting with the original fermions.

Bosonic system with constraints imposed is equivalent to the sector of the fermionic system (possibly including the spurious fermions) characterized by one of the two possible values of $(-1)^F$, defined including the spurious fermions. Which possibility is realized depends on the lattice geometry and the choice of values of $\gamma$ operators. In the remainder of this section we sketch the proof of this fact, while details have been given in \cite{BR2020} in a~slightly less general context.

First, on the space of solutions of the constraints all relations satisfied by $S$ and $T$ operators are obeyed by $\widehat S$ and $\widehat T$. Therefore this space is a representation of the algebra of even operators. Every representation of this algebra is a direct sum of irreducible representations, which are the two halves of the Fock space described by two values of $(-1)^F$. We will argue that only one of the two irreducible representations actually occurs in the decomposition and that the multiplicity is equal to one.

For the first part of the claim, it is sufficient to observe that the product of all $S$ and $T$ operators is proportional to the fermionic parity operator, while the product of all $\widehat S$ and $\widehat T$ is proportional to the product of all gamma matrices. The latter is proportional to $1$, because for every lattice site we have $\gamma(x)=1$ or $\gamma(x)=-1$. Combining these two results we conclude that $(-1)^F$ is represented by a scalar operator in the bosonic model. It is possible to determine whether it is equal to $+1$ or $-1$ by tracking phases carefully in the above argument. Details depend on the lattice geometry. 

For the second part of the claim it suffices to calculate the dimension of the space of solutions of constraints. This is facilitated by considering also modified forms of constraints. Those are in one-to-one correspondence with gauge orbits of background $\mathbb Z_2$ gauge fields. Thus there are $2^{N_1 - N_0 +1}$ of them, where $N_0$ is the number of lattice sites and $N_1$ is the number of edges. One can show that spaces of solutions of modified constraints all have the same dimension by explicitly constructing unitary operators which map between them. Therefore the dimension of the space of solutions of constraints \eqref{eq:constraints} is equal to the dimension of the whole Hilbert space in the bosonic model divided by $2^{N_1- N_0 +1}$. Using the well-known values of dimensions of irreducible representations of Clifford algebras we find that there are $2^{\frac{n}{2}-1}$ linearly independent solutions of constraints. This number is equal to the dimension of one half of the Fock space. 

\section{Examples} \label{sec:example}

We will now show how the general construction presented in the previous section works in specific examples. We begin with a model defined on the honeycomb lattice and discuss its relation with the Kitaev's model \cite{Kitaev06}. Then we discuss its three-dimensional deformation, the decagonal lattice. Discussion of boundary effects is postponed to Section \ref{boundary}.

We stress that this choice of lattices has been made mostly in order to simplify the presentation. Bosonization prescription from Sec. \ref{sec:1} is valid also on lattices of more complicated geometry: coordination numbers may vary from site to site and the translation symmetry is not necessary. We remark that our bosonization reduces in $(1+1)$-dimensions to the standard Jordan-Wigner transformation and as such can be thought of as its higher dimensional generalization.

\subsection{Honeycomb lattice and Kitaev's model}
\label{sec:hex}

\begin{figure}[htp]
\centering
    \begin{subfigure}[c]{0.15\textwidth}
         \centering
         \includegraphics[width=\textwidth]{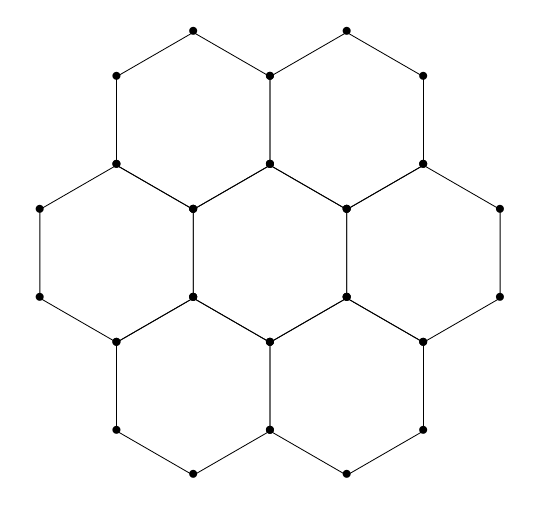}
         \caption{}
         \label{fig:00}
     \end{subfigure}
     \begin{subfigure}[c]{0.3\textwidth}
         \centering
         \includegraphics[width=\textwidth]{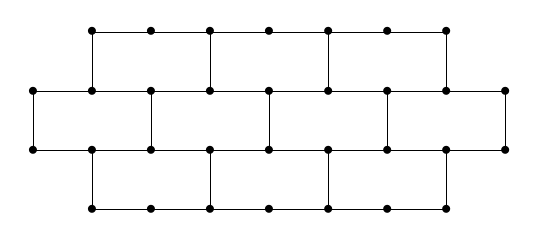}
         \caption{}
         \label{fig:02}
     \end{subfigure}
        \caption{(a) Honeycomb and (b) brick wall lattices are geometrically different, but topologically equivalent.}
        \label{fig:hex00}
\end{figure}

In this subsection we present an example involving a honeycomb lattice. It is arguably the simplest two dimensional lattice with vertices of odd degree. Since our construction is sensitive only to the topology rather than geometry of the lattice, honeycomb lattice is equivalent to the brick wall lattice, see Fig. \ref{fig:hex00}.  

\begin{figure}[htp]
    \centering
    \includegraphics[width=0.25\textwidth]{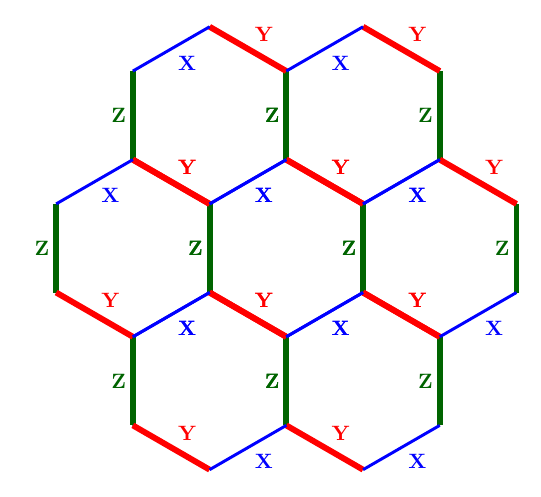}
    \caption{The assignment of Pauli matrices.}
    \label{fig:Kitaev}
\end{figure}

We will bosonize a system featuring one Majorana fermion $\psi$ per lattice site. This requires three gamma matrices. They can be represented by Pauli matrices $\sigma_X, \sigma_Y, \sigma_Z$, which are assigned to edges of the lattice as illustrated in Fig. \ref{fig:Kitaev}. Thus for a lattice site $x$ and an edge $e$ labeled by $I \in \{ X, Y, Z \}$ we have $\Gamma(x,e) = \sigma_I(x)$.

\begin{figure}[htp]
    \centering
    \includegraphics[width=0.2\textwidth]{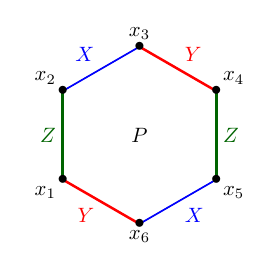}
    \caption{A plaquette $P$ of the honeycomb lattice.}
    \label{fig:miodek_jeden}
\end{figure}

For a plaquette $P$ depicted in Fig. \ref{fig:miodek_jeden}, the corresponding constraint takes the form $W_P | \mathrm{phys} \rangle = - | \mathrm{phys} \rangle$, where
\begin{equation}
W_P = \sigma_X(x_1) \sigma_Y (x_2) \sigma_Z(x_3) \sigma_X(x_4) \sigma_Y(x_5) \sigma_Z (x_6).
\end{equation}
Those are the Kitaev's plaquette operators \cite{Kitaev06}.

As a specific example, let us consider the Hamiltonian
\begin{equation}
H = i \sum\limits_{I \in \{ X, Y, Z \}}  \sum_{\substack{\text{type } I \\ \text{edges}}} J_I \, \psi(x) \psi(y),
\label{eq:Hf}
\end{equation}
where $x,y$ are the endpoints of the given edge, and $J_X, J_Y$ and $J_Z$ are parameters of the model. According to the prescription given in Eq. \eqref{eq:bosonization}, it corresponds to the spin Hamiltonian
\begin{equation}
\widehat{H} = - \sum\limits_{I \in \{ X, Y, Z \}}  \sum_{\substack{\text{type } I \\ \text{edges}}} J_I \, \sigma_I(x) \sigma_I(y),
\label{eq:Hb}
\end{equation}
subject to the constraint $W_P = -1$ for every plaquette $P$. This Hamiltonian has been proposed in \cite{Kitaev06}, where its study was reduced to diagonalization of quadratic fermionic Hamiltonians. Our approach provides an alternative derivation of this result. Subspaces defined by different values of $W_P$ correspond to the Hamiltonian $H$ modified by including a background $\mathbb Z_2$ gauge field. 

\subsection{Decagonal lattice} 

\begin{figure}[htp]
\centering
    \includegraphics[width=0.48\textwidth]{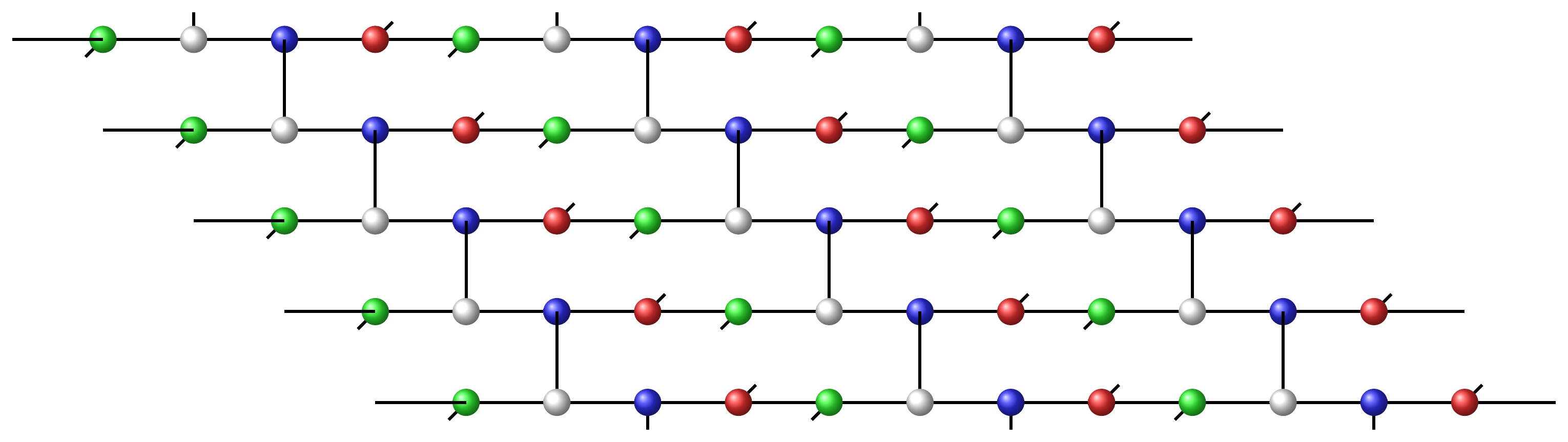}
    \caption{One layer of the decagonal lattice.}
    \label{fig:dec_one}
\end{figure}

An example of a three-dimensional trivalent lattice is provided by the decagonal geometry. A convenient representation is shown in Fig. \ref{fig:dec_one}, where one layer of such lattice is presented, together with edges connecting it with the adjacent layers. It can be thought of as a deformation of the brick wall lattice. In the brick wall geometry, each red site was connected with a green one to its north, while in the decagonal geometry it is instead connected with a green site lying in the layer underneath.

\begin{figure}[htp]
   \centering
   \begin{subfigure}[c]{0.3\textwidth}
         \centering
         \includegraphics[width=\textwidth]{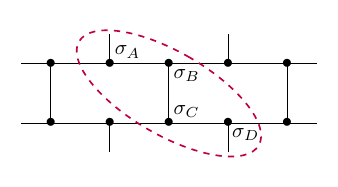}
         \caption{}
         \label{fig:hd_a}
     \end{subfigure}
     \begin{subfigure}[c]{0.3\textwidth}
         \centering
         \includegraphics[width=\textwidth]{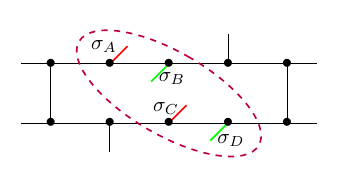}
         \caption{}
         \label{fig:hd_bb}
     \end{subfigure}
      \caption{The assignment of Pauli matrices for (a) the brick wall lattice and (b) the decagonal lattice.}
    \label{fig:hd}
\end{figure}

By the similarity with the brick wall geometry, one can easily generalize the results from Section \ref{sec:hex}. Using the identification of edges between honeycomb and brick wall lattices we attach Pauli matrices to pairs $(x,e)$ of the decagonal lattice, see Fig. \ref{fig:hd}. 

\begin{figure}[htp]
\centering
    \includegraphics[width=0.45\textwidth]{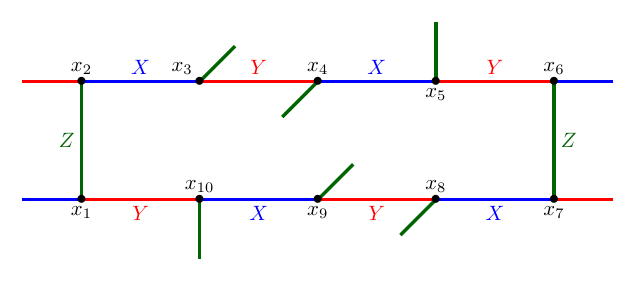}
    \caption{Plaquette $P$ within a single layer of the decagonal lattice.}
    \label{fig:dec_plaq}
\end{figure}

As an example, constraint associated to the plaquette from Fig. \ref{fig:dec_plaq} takes the form $W_P | \mathrm{phys} \rangle = - | \mathrm{phys} \rangle $, where
\begin{eqnarray}
    W_P  = &&  \sigma_X(x_1) \sigma_Y(x_2) \sigma_Z (x_3) \sigma_Z(x_4) \sigma_Z(x_5) \nonumber \\
      && \times  \sigma_X(x_6) \sigma_Y(x_7) \sigma_Z (x_8) \sigma_Z(x_9) \sigma_Z(x_{10}).
\end{eqnarray}
There exist also plaquettes not contained within one layer, but for the sake of brevity we will not write down the explicit formulas.
   
As in the honeycomb lattice case, every edge is labeled by $I \in \{ X, Y, Z \}$ and there is a correspondence between the Hamiltonians in Eqs. \eqref{eq:Hf} and \eqref{eq:Hb}.

\subsection{Hubbard model}

In the preceding examples only one kind of fermionic variables was involved. Here we discuss the simplest model with an additional quantum number involved - the Hubbard model \cite{hubbard} on the square lattice.

The Hamiltonian of the Hubbard model consists of two terms, $H=H_0+V$, where
\begin{subequations}
\begin{gather}
    H_0=-t\sum\limits_{\langle xy\rangle}\sum\limits_{\sigma=\uparrow,\downarrow}\left(c_{\sigma}^\dagger(x) c_{\sigma}(y)+c^\dagger_{\sigma}(y)c_{\sigma}(x)\right),
    \label{hubb0} \\
    V=U\sum\limits_x n_{\uparrow}(x)n_{\downarrow}(x),
    \label{hubbV}
\end{gather}
\end{subequations}
where the edge connecting sites $x$ and $y$ is denoted by $\langle xy \rangle$. Here $c_{\sigma}^\dagger(x)$ creates a fermion with spin $\sigma$ at position $x$, $n_{\sigma}(x)=c_{\sigma}^\dagger(x) c_{\sigma}(x)$,  and $t,U\in\mathbb{R}$. These fermionic operators may be decomposed into Majoranas as
\begin{subequations}
\begin{align}
    c_\uparrow(x)&=2^{-1}(\psi_0(x)+i\psi_1(x)),\\
    c^\dagger_\uparrow(x)&=2^{-1}(\psi_0(x)-i\psi_1(x)), \\
  c_\downarrow(x)&=2^{-1}(\psi_2(x)+i\psi_3(x)),  \\
    c^\dagger_\downarrow(x)&=2^{-1}(\psi_2(x)-i\psi_3(x)).
\end{align}
\end{subequations}
This choice is by no means unique and we are free to (consistently) use any other relabelling of indices. After modifying accordingly the bosonization prescription, different choices will lead to equivalent bosonic models.

To perform bosonization we need seven $\Gamma$ matrices per site, six of which are independent: the seventh may be taken to be the product of the first six and the imaginary unit. The bosonized Hamiltonian takes the form:
\begin{widetext}
\begin{subequations}
\begin{gather}
    \widehat H_0=-\frac{it}{2}\sum\limits_{e=\langle xy\rangle}\Gamma(x,e)\Gamma(y,e)\left(\Gamma_1'(x)-\Gamma_1'(y)-i\Gamma_2'(x)\Gamma_3'(y)+i\Gamma_3'(x)\Gamma_2'(y)\right), \\
    \widehat V=\frac{U}{4}\sum\limits_x\left(1-\Gamma_1'(x)\right)\left(1+i\Gamma_2'(x)\Gamma_3'(x)\right).
\end{gather}
\end{subequations}
\end{widetext}

\begin{figure}[htp]
\centering
    \includegraphics[width=0.2\textwidth]{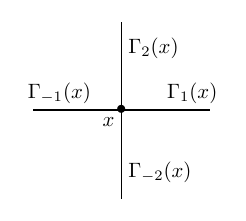}
    \caption{The assignment of gamma matrices.}
    \label{fig:sq_orient}
\end{figure}

\begin{figure}[htp]
\centering
    \includegraphics[width=0.2\textwidth]{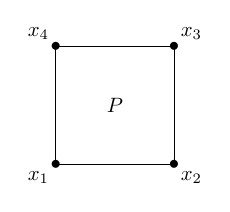}
    \caption{A plaquette $p$ of the rectangular lattice.}
    \label{fig:sq_verices}
\end{figure}

To write down the constraints, it is covenient to denote $\Gamma(x,e)$ as $\Gamma_i(x)$ for edge $i$ pointing from $x$ in the $i$-th direction, $i \in \{ \pm 1, \pm 2 \}$ (see Figure \ref{fig:sq_orient}). Now consider a~plaquette $P$ as in Fig. \ref{fig:sq_verices}. The corresponding constraint takes the form $W_P | \mathrm{phys} \rangle = - | \mathrm{phys} \rangle$, where
\begin{equation}
    W_P=\Gamma_{1,2}(x_1)\Gamma_{2,-1}(x_2)\Gamma_{-1,-2}(x_3)\Gamma_{-2,1}(x_4).
\end{equation}
Here we abbreviated $\Gamma_{i,j}(x) :=\Gamma_i(x)\Gamma_j(x)$. We note that this constraint does not at all involve primed gamma matrices, which we had to introduce in order to implement multiple fermions per site. It is characteristic for the square lattice geometry. 

One annoying feature of the presented construction is that spin up and spin down states are not treated completely symmetrically. Nevertheless, symmetries of the Hubbard model are implemented also in the bosonized model. First, we have conservation of the total particle number. Particle number on a single lattice site $x$ is bosonized in the following way:
\begin{equation}
    \sum_{\sigma} c_{\sigma}^{\dagger} c_{\sigma} \longleftrightarrow 1 - \frac12 \Gamma_1'+ \frac{i}{2} \Gamma_2' \Gamma_3'.
\end{equation}
For spin ($\mathrm{SU}(2)$ generators) operators we have:
\begin{subequations}
\begin{gather}
c_{\uparrow}^\dagger c_{\uparrow} - c_{\downarrow}^\dagger c_{\downarrow} \longleftrightarrow - \frac12 \Gamma_1' - \frac{i}{2} \Gamma_2' \Gamma_3', \\
c_{\uparrow}^{\dagger} c_{\downarrow} \longleftrightarrow \frac{i}{4} (1 - \Gamma_1')(\Gamma_2' + i \Gamma_3'), \\
c_{\downarrow}^\dagger c_{\uparrow} \longleftrightarrow - \frac{i}{4} (1 + \Gamma_1')(\Gamma_2' - i \Gamma_3').
\end{gather}
\end{subequations}
It noteworthy that all on-site symmetries of the original fermionic model are also on-site after bosonization. Furthermore, the corresponding charges are expressed entirely in terms of \emph{primed} gamma matrices. This is a~general property of our construction. 

\section{Boundary effects}\label{boundary}

\subsection{Rectangular lattice with a boundary}

We will now discuss bosonization of a system on an $L_x \times L_y$ rectangular lattice, with two Majorana fermions $\psi_0$, $\psi_1$ per lattice site. Every site $x$ in the bulk has four neighbors, corresponding to four gamma matrices $\Gamma_{\pm i}(x)$, $i=1,2$, as in our discussion of the Hubbard model. According to the prescription given in Sec. \ref{sec:1}, we need also an additional gamma matrix $\Gamma'_1(x)$. It can be eliminated by imposing relations discussed below equation \eqref{eq:gamma}. We choose the convention $\Gamma_1'(x) = \Gamma_{-1}(x) \Gamma_1(x) \Gamma_{-2}(x) \Gamma_2(x)$. Therefore in the end we need only unprimed gamma matrices. Constraints are identical as in the discussion of the Hubbard model. 

More explicitly, our bosonization prescription reads
\begin{itemize}
    \item $i \Gamma_1(x) \Gamma_{-1}(y) \longleftrightarrow  \psi_0(x)\psi_0(y)$ if $y$ is the eastern neighbor of $x$, 
    \item $i \Gamma_2(x) \Gamma_{-2}(y) \longleftrightarrow  \psi_0(x)\psi_0(y)$ if $y$ is the northern neighbor of $x$,
    \item $i \Gamma_{-1}(x) \Gamma_1(x) \Gamma_{-2}(x) \Gamma_2(x) \longleftrightarrow \psi_0(x) \psi_1(x)$.
\end{itemize}

\begin{figure}[htp]
\centering
    \includegraphics[width=0.35\textwidth]{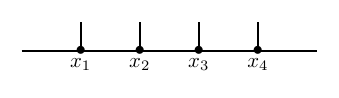}
    \caption{The southern boundary of the rectangular lattice.}
    \label{fig:south_border}
\end{figure}

Now, we look closely at the situation on the boundary. First, sites on the southern edge (see Fig. \ref{fig:south_border}) have no neighbors in the direction $-2$. We may reinterpret the $\Gamma_{-2}$ matrix as an additional $\Gamma'$, corresponding to a spurious Majorana fermion on the boundary. More precisely, for every site $x_i$ on the southern edge we introduce an additional Majorana operator $\chi_{\mathrm S}(x_i)$. Bosonization prescription for $\chi_S$ fermions takes the form 
\begin{equation}
    i\Gamma_{-2}(x_i) \longleftrightarrow \psi_0(x_i)\chi_{\mathrm S}(x_i).
    \label{eq:chiS_boso}
\end{equation}
Similarly for the northern, eastern and western edges we introduce Majorana fermions $\chi_{\mathrm N}$, $\chi_{\mathrm E}$ and $\chi_{\mathrm W}$.

At each of the four corners (which are geometrically of codimension two) there are two $\chi$ fermions. For example the south-east corner $x_{\mathrm{SE}}$ hosts four Majorana operators $\psi_0(x_{\mathrm{SE}})$, $\psi_{1}(x_{\mathrm{SE}})$, $\chi_S(x_{\mathrm{SE}})$ and $\chi_E(x_{\mathrm{SE}})$.

We now determine the identity resulting from existence of the boundary. First, notice that for every lattice site $x=(a,b) \in \{ 1, \ldots , L_x \} \times \{ 1 , \ldots, L_y \}$ we have
\begin{equation}
    \prod\limits_{a=1}^{L_x-1}\psi_0(a,b)\psi_0(a+1,b)=\psi_0(a,b)\psi_0(L_x,b).
\end{equation}
Consequently, our bosonization prescription yields
\begin{equation}
\begin{split}
   & \psi_0(1,b)\psi_0(L_x,b)\\ &\longleftrightarrow i^{L_x-1}\Gamma_{-1}(1,b)\left(\prod\limits_{a=1}^{L_x}\Gamma_{-1,1}(a,b)\right)\Gamma_1(L_x,b)
    \end{split}
\end{equation}
and
\begin{equation}
\begin{split}
    &\psi_0(a,1)\psi_0(a,L_y)\\
    &\longleftrightarrow i^{L_y-1}\Gamma_{-2}(a,1)\left(\prod\limits_{b=1}^{L_y}\Gamma_{-2,2}(a,b)\right)\Gamma_1(a,L_y),
\end{split}
\end{equation}
for every $1\le b\le L_y$ and $1\le a\le L_x$, respectively. By a~straightforward computation one can check that it results in the following correspondence
\begin{equation}
\begin{split}
    &i^{(L_x-L_y)^2+2(L_x+L_y)}\prod\limits_{a,b}\Gamma_{-2,2,-1,1}(a,b)\\
    &\longleftrightarrow \chi_{\partial_S}\chi_{\partial_N}\chi_{\partial_W}\chi_{\partial_E},
\end{split}
\end{equation}
where we have introduced the abbreviated notation $\chi_{\partial_S}=\prod\limits_{a=1}^{L_x}\chi_S(a,1)$, and so on. Since $\Gamma_{-2,2,-1,1}(x)$ corresponds to $-i\psi_0(x)\psi_1(x)$, which is the parity operator $(-1)^{F_\psi(x)}$ for $\psi$ fermions at site $x$, we end up with the following constraint
\begin{equation}
    (-1)^{F_{\psi}}= \kappa  \chi_{\partial_S}\chi_{\partial_N}\chi_{\partial_W}\chi_{\partial_E},
\end{equation}
where $\kappa=i^{-(L_x-L_y)^2+2(L_x+L_y)}$ is a geometrical phase factor. In particular, for lattices with $L_x\equiv L_y\pmod{2}$ we have $\kappa=1$.

Summarizing, we started from the system of $\psi$ fermions, but our bosonization gave us a bosonic system equivalent to $\psi$ fermions together with $\chi$ fermions on the boundary. Since operators corresponding on the spin side to spurious $\chi$ modes have been identified, this is not a problem. Indeed, bosonizing suitable Hamiltonian for $\chi$ fermions using formula \eqref{eq:chiS_boso} and its analogues for other components of the boundary we may make $\chi$ arbitrarily heavy, e.g. dimerized with large dissociation energy. This provides a physical interpretation for additional constraints on the boundary introduced in \cite[Appendix~B]{BR2020}. After imposing them, one obtains a spin system corresponding to $\psi$ fermions on a lattice with boundary.

\subsection{Comment about topological phases}

One of the remarkable features of many topologically nontrivial phases of matter is their interesting (robust)  behaviour on the boundaries, which are of dimension $d-1$. This is the case in particular for free fermion systems, for which one has a well-established bulk-boundary correspondence: topological invariants in the bulk signal existence of modes localized near the boundary, responsible for closing the gap in finite volume. 

Robustness of the boundary modes in often interpreted as manifestation of an anomaly of the boundary theory. Presence of the anomaly implies existence of some degrees of freedom ``saturating'' the anomaly. On the other hand, it is also expected that the anomalous $(d-1)$-dimensional system is inconsistent on its own: it may be realized only on the boundary of a $d$-dimensional system. As an example, chiral fermions cannot be realized on the lattice (this is the Nielsen-Ninomiya theorem, see \cite{NIELSEN198120, Friedan}), but they may exist on the boundary or domain wall (more generally, a defect) in a higher dimensional system. This is at the heart of the bulk boundary correspondence.

On the other hand, boundary modes found in our bosonization prescription correspond to a standalone (hence ``non-anomalous'') system on the boundary. Now suppose that we bosonize a free fermion system with a nonzero topological invariant, say on a half-space. Then on the boundary we will have boundary modes predicted by the bulk-boundary correspondence and $\chi$ fermions described in the previous subsection. We can gap out the latter fermions by including in the Hamiltonian a suitable term localized in the boundary. This is believed not to be true for the former. In the case of invariants which remain robust in presence of interactions \cite{Kitaev_wire} it is natural to expect that even after including a coupling between $\chi$ fermions and $\psi$ fermionis, boundary modes originating from a topological invariant will persist. 

Of course the full picture of topological invariants and boundary modes has to involve the choice of a Hamiltonian, or at least some class of Hamiltonians. On the other hand, the discussion presented here is mostly concerned with properties of algebras of observables. It would be interesting to understand better the relation between bosonization and bulk-boundary correspondence. Such questions are relevant, for example, for the problem of discretization of chiral fermions. 

\section{Euclidean representation of unconstrained
``Majorana spins"} \label{Euclid}

The next goal is to construct an Euclidean Ising-like action, with two different couplings, $\beta_t$ and $\beta_s$, which in the continuous time limit  
\begin{equation}
    \beta_t\rightarrow\infty,\quad \epsilon=e^{-\beta_t}\rightarrow 0,\quad \beta_s=\epsilon\lambda \rightarrow 0
    \label{ctl}
\end{equation}
is described by the Hamiltonian \eqref{eq:Hb} with parameters $J_X=J_Y=1$, $J_Z = \lambda$ (we will not impose constraints at this point). Following \cite{FS78,KOGUT}, this is done by demanding that the transfer matrix elements defined by  the Boltzmann weight
\begin{equation} 
\langle s ' |  T | s \rangle=e^{-\mathsf L(s',s)} 
\end{equation}
 coincide with these of the Euclidean evolution operator
\begin{equation}
\langle s ' |  T | s \rangle= \langle s ' |  e^{- \epsilon H} | s \rangle = \langle s ' |  1 - \epsilon H | s \rangle
\end{equation}
up to terms of order $\epsilon$. Here $\epsilon$ is the elementary time step and $s$ and $s'$ denote configurations of spins at subsequent time instants.

The time evolution generated by \eqref{eq:Hb} consists of elementary double-spin flips, in contrast to the Ising system in which the dynamics is driven by single spin flips. In order to gain some orientation in this problem, we start by deriving an Euclidean action for a simpler, one-dimensional quantum Hamiltonian
\begin{equation}
\begin{split}
H_{1d}=&-\sum_k \sigma_X(x_k)\sigma_X(x_{k+1}) \\
&- \lambda \sum_k \sigma_Z(x_k)\sigma_Z(x_{k+1}).  \label{H1d}
\end{split}
\end{equation}

\subsection{Basic idea and the $(1+1)$-dimensional example}
In the Ising model, the basic trick relating Hamiltonian and functional formulations is to classify all variations of a multiple-spin state into classes with fixed number of single spin flips. 

On the Euclidean side, the number of single flips between two time slices is counted by the two-row action
\begin{equation}
\mathsf L_1(s',s)=\frac{1}{2}\sum_k (1-s_ks_k').   
\end{equation}
It corresponds to the Hamiltonian 
\begin{equation}
   - \sum_k \sigma_X(x_k).
\end{equation}

In the present case \eqref{H1d}, we are seeking to single out the {\em double spin flips} out of all possible changes of a row of spins. Therefore we begin with the Euclidean eight-spin action which counts {\em isolated} double flips 
\begin{equation}
\begin{split}
\mathsf L_2^{(8)}&=\frac{1}{2^4}\sum_k (1+s_{k-1} s_{k-1}')(1-s_{k}s_{k}')\\
&\times
(1-s_{k+1} s_{k+1}')(1+s_{k+2} s_{k+2}'), \label{S2V}  
\end{split}
\end{equation}
Simpler functions can be also used, hence we shall omit the ``$(8)$" superscript if not necessary.

We need to arrange the final, Euclidean action such that in the continuous time limit it gives weight $\epsilon$ to double flips while all other, single and multiple, flips are of higher order in $\epsilon=e^{-\beta_t}$.
This is achieved by the combination
\begin{equation}
\mathsf L_{\mathrm{kin}}(s',s)=\beta_t \left( p(\mathsf L_1-2 \mathsf L_2) + \mathsf L_2 \right), \label{ST}
\end{equation}
where $p \geq 2$ is a free parameter. 

It is easy to check that $\mathsf L_2^{(8)}$ may also be replaced in \eqref{ST} by the simpler function 
 \begin{equation}
\mathsf L_2^{(6)}=\frac{1}{8}\sum_k (1 + s_{k-1} s_{k-1}')(1-s_{k} s_{k}')(1-s_{k+1} s_{k+1}'). \label{S1V}     
 \end{equation} 
This definition prescribes different weights to non-leading transitions, but results in the same double flip kinetic part of \eqref{H1d}. This is an illustration of the well known fact that many different Euclidean discretizations have the same continuous time limit, hence also the same Hamiltonian. 

Action for a single transition has to be supplemented by a potential term:
\begin{subequations}
\begin{gather}
    \mathsf L(s',s) = \mathsf L_{\mathrm{kin}}(s',s) + \mathsf L_{\mathrm{pot}}(s',s), \\
    \mathsf L_{\mathrm{pot}}(s',s) = - \frac{\beta_s}{2} \sum_k \left( s_k s_{k+1} + s_{k}' s_{k+1}' \right).
    \label{eq:Ltot}
\end{gather}
\end{subequations}

Complete Euclidean action for the $L_x \times L_t$ spins is obtained by composing elementary transfer matrices, which amounts to adding the corresponding actions:
\begin{equation}
S(s(L_t), \ldots , s(1))= \sum\limits_{t=1}^{L_t} \mathsf L (s(t+1),s(t)) \label{Spr1d}.
\end{equation}

This concludes our construction of the two-dimensional, Euclidean system which in the continuum time limit is described by the Hamiltonian \eqref{H1d}.

\subsubsection{$\sigma_Y\sigma_Y$ terms - the phases.}
The second example deals with the phase generating kinetic terms
\begin{align}
H_{1d}^{\mathrm{ph}}=&-\sum_{k\; \mathrm{even}}\sigma_X(x_k)\sigma_X(x_{k+1}) -\sum_{k\; \mathrm{odd}}\sigma_Y(x_k)\sigma_Y(x_{k+1}) \nonumber \\
&- \lambda \sum_k \sigma_Z(x_k)\sigma_Z(x_{k+1})  \label{H12}   
\end{align}
still in one space dimension.

Begin with an evolution of a two spin system: 
\begin{equation}
 s=\{s_1,s_2\}\rightarrow s'=\{s_1',s_2'\}.
\end{equation} As far as the change of spin states is considered, the action of $\sigma_Y\sigma_Y$ is the same as that of $\sigma_X\sigma_X$. The only difference is a phase factor:
\begin{equation}
\begin{split}
&\sigma_Y(x_1)\sigma_Y(x_2) | s_1,s_2 \rangle  \\ &= \exp{\left(  \frac{i \pi}{2}(s_1+s_2)\right)}  \sigma_X(x_1)\sigma_X(x_2) | s_1,s_2 \rangle.
\label{phase}
\end{split}
\end{equation}
Generalization to a whole row of $L$ spins is straightforward. The kinetic term of the Hamiltonian \eqref{H12} will be reproduced by the action \eqref{ST} supplemented by a phase \eqref{phase} for each odd edge. This gives for the new action of the two complete rows (and with the unchanged diagonal potential term)
\begin{widetext}
\begin{equation}
\begin{split}
\mathsf{L}_{\mathrm{ph}}(s',s)&=\beta_t  \left(2(S_1-2 S_2) + S_2 \right)+ \beta_s S^{\mathrm{pot}}\\
&+\frac{i \pi}{2} \frac{1}{2^3}\sum_{x - \mathrm{odd}} (s_x+s_{x+1}) (1+s_{x-1} s_{x-1}')(1-s_{x}s_{x}')(1-s_{x+1}s_{x+1}'). 
\end{split}
\label{L2ph}
\end{equation}
\end{widetext}

\subsection{$(2+1)$-dimensional system}
As a $(2+1)$-dimensional example we consider here the honeycomb lattice discussed in \ref{sec:hex}. As remarked therein, it is convenient to represent it in a brick wall form, cf. Fig. \ref{fig:hex00}.

\begin{figure}[htp]
\centering
    \includegraphics[width=0.45\textwidth]{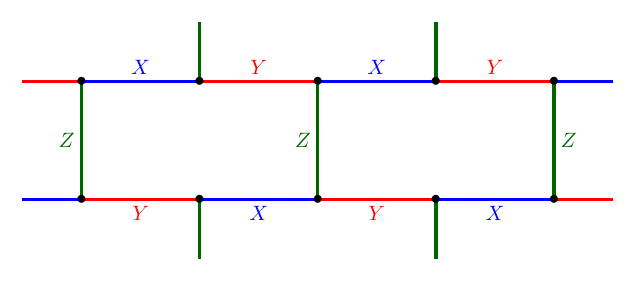}
    \caption{The brick wall lattice with the assignment of Pauli matrices.}
    \label{fig:hbi}
\end{figure}

Our Hamiltionian is of the form \eqref{eq:Hb} with parameters $J_X=J_Y=1$ and $J_Z=\lambda$, and can be written as a sum of two terms
\begin{equation}
    H=H_{\mathrm{kin}}+\lambda H_{\mathrm{pot}}, \label{H3D}
\end{equation}
where the kinetic term contains sums over all edges of type $X$ and $Y$, while the potential one is a  sum over edges of type $Z$. The labelling of the edges of the brick wall lattice is shown explicitly in Fig. \ref{fig:hbi} and is consistent with the one in Fig. \ref{fig:Kitaev}.

Derivation of an Euclidean action of a three-dimensional $(x,y,t)$, periodic in all directions, system is very similar to the previous $(1+1)$-dimensional example. 

The two kinetic (i.e. $\sigma_X\sigma_X$ and  $\sigma_Y\sigma_Y$) terms in \eqref{H3D} are represented by the same six- or eight-spin couplings between the adjacent time slices plus the appropriate phase, which naturally generalizes
the $(1+1)$-dimensional phase in the last term of \eqref{L2ph} to three Euclidean dimensions.

On the other hand diagonal, in the Hamiltonian form, potential terms are represented by the standard Ising-like, ferromagnetic couplings along the $y$-direction. They are located on the shorter edges of bricks at each time slice. Hence, they are staggered in  accord with the ($t$-independent) $x-y$ parity, $\zeta_{xy}=(-1)^{x+y}$, of a site originating given $Z$-edge in the potential term. The final action reads
\begin{equation}
\begin{aligned}
    S_{3D}&=\beta_t \sum_{x,y,t} O^{(6)}_{x,y,t}+\beta_s \sum_{\substack{x,y,t,\\ \zeta_{xy}=1}} O^{(2)}_{x,y,t}\\
    &+ \frac{i \pi}{2}\sum_{\substack{x,y,t,\\\zeta_{xy}=-1}}O^{(7)}_{x,y,t}, \label{S3D}
\end{aligned}
\end{equation}
with the phase operator $O^{(7)}_{x,y,t}$ being the direct generalization of  above $O^{(7)}_{x,t}$ to three dimensions and similarly for other couplings:
\begin{widetext}
\begin{subequations}
\begin{gather}
    O^{(7)}_{x,y,t} = \frac{1}{2^3}(s_{x,y,t}+s_{x+1,y,t}) (1+s_{x-1,y,t} s_{x-1,y,t+1})(1-s_{x,y,t}s_{x,y,t+1})(1-s_{x+1,y,t}s_{x+1,y,t+1}), \\
    O^{(6)}_{x,y,t}= \frac{1-2 p}{8} (1+s_{x-1,y,t} s_{x-1,y,t+1})(1- s_{x,y,t} s_{x,y,t+1})(1- s_{x+1,y,t} s_{x+1,y,t+1}) + \frac{p}{2}(1- s_{x,y,t} s_{x,y,t+1}), \\
    O^{(2)}_{x,y,t}= - s_{x,y,t}s_{x,y+1,t}.
\end{gather}
\end{subequations}
\end{widetext}

The action \eqref{S3D} describes then a three-dimensional Ising-like system. Together with the corresponding constraints (still to be implemented) it would provide an equivalent, Euclidean representation of a single, quantum Majorana spin on a two-dimensional spatial lattice.   

Even without the constraints the system is still interesting {\it per se}. Its thermodynamics, the phase diagram, order parameters are unknown at the moment and could be studied with standard methods of statistical physics. Such studies would also provide, among other things, some information about the constraints themselves.

The Boltzmann factor associated with  \eqref{S3D} is not positive. However the origin of its phases is now conceptually simple. Below we look how severe is the sign problem in these unconstrained Euclidean models.

\subsection{The sign problem}
The standard (and practically only) method to deal with non-positive weights is the reweighting \cite{reweight,nakamura}. Instead of potentially negative Boltzmann factor $\rho=\exp{(-S)}$, one uses as a Monte Carlo (MC) weight its absolute value $\rho_A=|\rho|$, correcting at the same time all observables for this bias.

Whether such an approach is practical can be readily judged from the average value of a sign of the exact Boltzmann factor 
\begin{equation}
    \langle \mathrm{sign}\rangle\equiv\left\langle\frac{\rho}{\rho_A}\right\rangle_A=\frac{\mathcal{Z}}{\mathcal{Z}_A}
\end{equation}
averaged over the modulus $\rho_A$. If this average is close to $0$, the method fails. If the contrary is true, say for some intermediate volumes, one may expect to obtain meaningful estimates.

We have calculated analytically  above average for both $(1+1)$- and $(2+1)$-dimensional models by employing the transfer matrix  technique for a range of small volumes. It is seen below that the sign problem is not very severe in this case. Consequently, MC studies remain a viable approach to explore these systems in detail.

\begin{figure*}[htb]
\includegraphics[width=\textwidth]{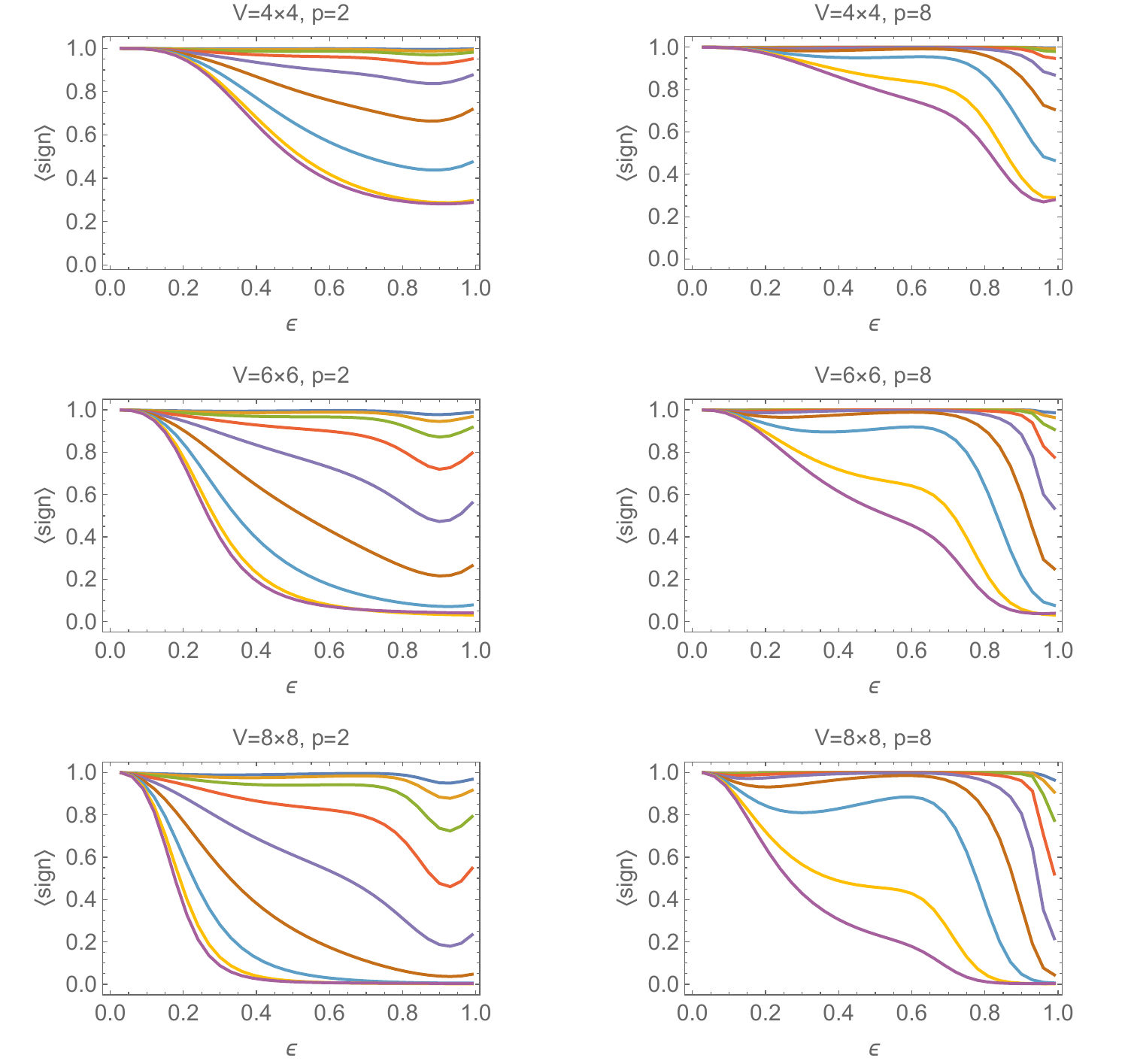}
\caption{Exact results for the average sign $\langle \mathrm{sign}\rangle$ for a range of two dimensional volumes $V$ and for the penalty parameter $p=2$ (left column) or $p=8$ (right column). Plots are presented for $\lambda$ values (from bottom to top) $0.1$, $0.25$, $0.5$, $0.75$, $1$, $1.25$, $1.5$, $1.75$ and $2$.} 
\label{fig:f1}
\end{figure*}

\subsubsection{$(1+1)$-dimensions}
Partition functions $\mathcal{Z}$ and $\mathcal{Z}_A$ were calculated exactly by summing Boltzmann factors $\exp{(-S_{2D})}$ and $|\exp{(-S_{2D})}|$, as defined in Eqns \eqref{L2ph}.
In Fig.\ref{fig:f1} the average sign is shown for a range of two dimensional volumes and various couplings $\beta_t$ and $\beta_s$. The results are displayed as a function of a time step, $\epsilon=\exp{(-\beta_t)}$, and parameterized by different couplings $\lambda=\frac{\beta_s}{\epsilon}$ in the Hamiltonian \eqref{H12}. Second column displays analogous results for larger penalty parameter $p$.

The sign problem seems manageable for a sizeable part of the parameter space. It vanishes entirely for $\epsilon \rightarrow 0$. 

Increasing the penalty parameter $p$ also helps, since then some undesired transitions vanish faster with $\epsilon$.

Both of these features show up also in our three-dimensional system. They can be readily understood and used for our advantage, as discussed below.

\begin{figure}[htb]
\centering
\begin{subfigure}[c]{0.45\textwidth}
         \centering
         \includegraphics[width=\textwidth]{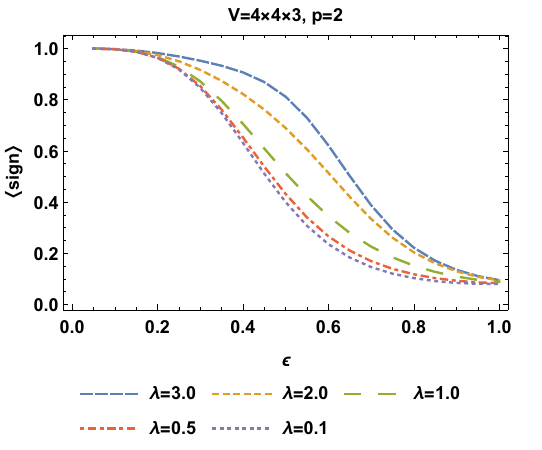}
         \caption{}
         \label{fig:f2_00}
     \end{subfigure}
     \begin{subfigure}[c]{0.45\textwidth}
         \centering
         \includegraphics[width=\textwidth]{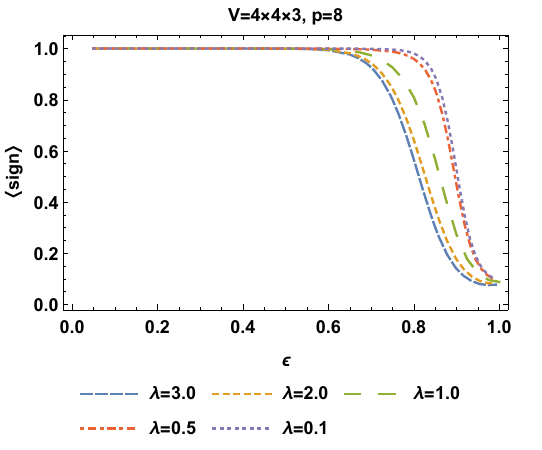}
         \caption{}
         \label{fig:f2_02}
     \end{subfigure}
\caption{Average sign as a function of the $\epsilon$ parameter in the three-dimensional case with volume $V=4\times 4\times 3$  for the penalty parameter (a) $p=2$ and (b) $p=8$. The range of $\lambda$ parameter is explicitly given for both cases.}
\label{fig:f2}
\end{figure}

\subsubsection{$(2+1)$-dimensions}
 For the three-dimensional Euclidean system \eqref{S3D} of volume $V=L_x L_y L_t$ a brute-force summation of all $2^V$ terms becomes already a challenge. Still it was possible to obtain the value of $\langle\mathrm{sign}\rangle$ for $V=4\times 4\times 3$, as shown in Fig. \ref{fig:f2}. It was done by constructing two subsequent transfer matrices in the $y$ direction. 

Again, as in the $(1+1)$-dimensions, the phase is harmless for small $\epsilon$. This feature improves dramatically with increasing the penalty parameter.

In addition, for $L_t=2$ no phase was observed in all cases. That is $\langle\mathrm{sign}\rangle=1$ for all values of parameters and for all studied dimensions.

\subsubsection{The sign problem - summary}
 
All the regularities observed above can be readily understood and generalized for arbitrary sizes of lattices, providing at the same time some guidelines for other, similar systems.

Consider first the case $L_t=2$. The partition function 
\begin{equation}
    \mathcal{Z}^{(2)}= \sum_{s,s'} e^{- \mathsf L(s,s')} e^{- \mathsf L(s',s)}
\end{equation}
is the sum over two-composite states of spins at the two time slices. The non-zero phase can occur only if $s$ and $s'$ differ by a double flip. However in this case the phases of $e^{- \mathsf L(s,s')}$ and $e^{- \mathsf L(s',s)}$ cancel and the result is positive for each pair of configurations, as found above.

On the other hand, already for $L_t=3$ there are three states in the game
\begin{equation}
    \mathcal{Z}^{(3)}= \sum_{s,s',s''} e^{- \mathsf L(s,s'') - \mathsf L(s'',s') - \mathsf L(s',s)}.
\end{equation}
Hence a single double-flip, e.g. in $s \rightarrow s'$, can be balanced by two subsequent single-flips in $s' \rightarrow s''$ and $s'' \rightarrow s$ transitions. Since a phase may occur only in the double flip transition $s \rightarrow s'$, this particular contribution may be negative and would give $\langle\mathrm{sign}\rangle < 1$.

Consequently, the single flip transitions provide an undesired background which indirectly causes negative signs of Boltzmann factors, hence the sign problem.

However such transitions vanish for $\epsilon \rightarrow 0$ having a weight of the higher order in $\epsilon$ by construction. This is clearly confirmed by our calculations, cf. Figs \ref{fig:f1} and \ref{fig:f2}, and explains why sign problem vanishes at small $\epsilon$. These figures were obtained by using Mathematica \cite{MATH}.

Moreover, by increasing the penalty parameter $p$ we can force the ``bad transitions"  to vanish faster. Indeed this is also confirmed by our results for $p=8$ in both dimensions. This suggests that the sign problem could be significantly reduced by setting $p=\infty$, which amounts to introducing a constraint in the Euclidean system \footnote{Not to be confused with plaquette constraints required for bosonization.}. 

 We remark that even under this constraint there exist ``Euclidean histories'' with negative sign. As they involve a number of spin flips growing with the system size, one may hope that they do not lead to significant difficulties. 
 
 Obviously all these scenarios should be further studied quantitatively.

\section{Conclusions and outlook}

We have presented a bosonization method generalizing the idea from \cite{Wosiek:1981mn}, valid for lattices of arbitrary coordination number and with arbitrary number of Majorana modes per lattice site. In the previous works only systems with even coordination numbers and one pair of fermionic creation/annihilation operators per lattice site were considered. The new approach extends the construction in several ways. First, for lattices with vertices of even degree we may include multiple fermionic states per site. We illustrate this by bosonizing the Hubbard model. Second, we allow for lattices with odd coordination numbers. Then there is an odd number of Majorana fermions per site. We stress that the Majorana variables we are talking about here are not necessary resulting from any representation of complex (Dirac) fermions, but they are the elementary objects per se. In particular systems with one Majorana per site may be bosonized. Since the presented bosonization procedure is clearly invertible (as it is based on an algebraic isomorphism), this leads to an intriguing possibility of analyzing other spin liquids by applying the inverse of it. We have illustrated this general phenomena on the simplest example, the Kitaev's honeycomb lattice, but one can apply this procedure to other models of this type. Similar constructions based on Clifford algebras formalism have been previously, as discussed in Sec.\ref{sec:intro}, considered in \cite{NS2009,Yao2009,Wu2009} in order to fermionize higher spin models. More recently the gamma-matrix versions of Kitaev's models were used to study spin-$\frac{3}{2}$ Kitaev Shastry-Sutherland model \cite{Eschmann20} as well as to describe spin-orbital models and they relations to Kugel-Khomskii-type models and compass interactions \cite{Chulliparambil20}. In the latter case the authors constructed models, on either rectangular or honeycomb lattices, realizing the Kitaev's sixteenfold way of anyons \cite{Kitaev06}. Our bosonization method provides a rigorous mathematical technique that, in principle, could be use to generalize such construction in other geometries. Three-dimensional Kitaev's spin liquids were also studied recently in \cite{Eschmann}. Several examples of possible use of $\Gamma$-Kitaev models to study higher spin models as well as spin-orbital models were also reported in \cite{Chern10}, and used in \cite{Wu2009,Ryu09} to demonstrate the existence of emergent topological insulators on a three-dimensional diamond lattice. Since our bosonization provides tools for a rigorous construction (out of almost arbitrary fermionic theories) of bosonic (higher spin) models in terms of gamma matrices, it can be also used to generate new examples of (higher) spin models. We postpone this intriguing possibility for a future research.

It is possible to treat also systems for which the coordination number is not congruent modulo two to the number of Majoranas per site. Strictly speaking in this case we do not bosonize the original fermionic system but rather one augumented by some spurious fermionic degrees of freedom. Nevertheless, operators corresponding on the bosonic side to these modes may be clearly identified and decoupled. Even in the case of very regular lattices such trick is needed in presence of a boundary. We emphasize that this is a feature of our bosonization method, not of fermionic systems per se. 

If these two numbers are not congruent modulo two, it involves spurious fermionic states, which nevertheless can be identified and eliminated. Another potential source of interest in this construction is that it provides new analytically tractable examples of spin systems featuring edge modes.

One question which remains unanswered is whether our construction may be dualized to some higher gauge theory. For systems with one fermion (and hence two Majoranas) per lattice site such picture of bosonization has been obtained in \cite{CHEN2018234,ChenKapustin19}.

Concerning the Euclidean formulation, our main conclusion is that in spite of somewhat unusual time evolution, generated by simultaneous double-flips, a local Euclidean action for an unconstrained system was derived. It contains at least six-spin interactions and is highly asymmetric between space and time, in contrast to the standard Ising model. To our knowledge, this system has not been studied. Now, it can be readily explored with standard statistical methods. 

Our generic action is complex. It was found that the resulting sign problem is manageable on small lattices where our fully analytical approach is available. 

The next logical step now is to study the problem for larger, although intermediate, sizes and see whether the popular reweighting methods allow meaningful measurements of observables, extrapolation to larger volumes and extraction of scaling limits. Moreover, it is conceivable  that introduced here methods could be extended to implement the spin constraints avoiding the standard non-positive Legendre transformation. We intend to further study some of these questions with the aid of quantitative Monte Carlo approach. 

\begin{acknowledgments}
The authors acknowledge W.~Brzezicki for suggesting the decagonal lattice as a three-dimensional example with odd degree vertices. AB acknowledges G.~Ortiz and Z.~Nussinov for pointing out the references \cite{NS2009, Yao2009, Wu2009} and for discussion about them. This work is supported in part by the NCN grant: UMO-2016/21/B/ST2/01492. BR was also supported by the MNS donation for PhD students and young scientists N17/MNS/000040.
\end{acknowledgments}

\bibliography{main.bib}

\end{document}